\documentclass[prl,twocolumn,showpacs,floatfix,showkeys]{revtex4-1}
\usepackage{times,amsmath,amssymb,amstext,latexsym,float,graphicx,color}
\usepackage[dvipsnames]{xcolor}

\usepackage[hidelinks]{hyperref}
\hypersetup{colorlinks=true, citecolor=blue, urlcolor=blue, linkcolor=blue}

\begin{document}

\title{Mixed bubbles in a one-dimensional Bose-Bose mixture}
\author{P. Stürmer}
\email[]{philipp.sturmer@matfys.lth.se}
\affiliation{Mathematical Physics and NanoLund, Lund University, Box 118, 22100 Lund, Sweden}
\author{M. Nilsson Tengstrand}
\affiliation{Mathematical Physics and NanoLund, Lund University, Box 118, 22100 Lund, Sweden}
\author{S. M. Reimann}
\affiliation{Mathematical Physics and NanoLund, Lund University, Box 118, 22100 Lund, Sweden}

\date{\today}
\begin{abstract}
We investigate a Bose-Bose mixture across the miscible-immiscible phase transition governed by quantum fluctuations in one dimension. We find the recently predicted so-called \emph{'mixed bubbles'} as ground states close to the mean-field miscible-immiscible threshold. These bubbles form a pocket of miscibility, separated by one of the components. The collective excitations reflect the symmetry breaking resulting from the bubble formation. The partial miscibility of the system allows for persistent currents in an annular confinement. Intriguingly, the mixed bubble acts like an intrinsic weak link, connecting the rotational behavior of the mixed bubble state to current efforts in atomtronic applications.
\end{abstract}

\maketitle
In the mean-field (MF) approximation the phase of a two-component Bose-Bose mixture is dominated by its interaction coupling strengths $g_{\sigma\sigma'}$ between the components $1$ and $2$ with particle densities $n_1$ and $n_2$. The system is determined by $\delta g = g_{12}-\sqrt{g_{11}g_{22}}$ and may be in a miscible phase if $\delta g < 0$ and in an immiscible phase if $\delta g > 0$~\cite{ref:bookPethickSmith,ref:bookStringari}. Experiments on Bose-Bose mixtures have shown the phase transition of repulsive mixtures from miscible to immiscible~\cite{ref:sepPapp,ref:spinKetterle}, in agreement with theoretical predictions~\cite{ref:HFEsry,ref:PhaseSepTimmermans}, and the formation of dark/bright solitons was also shown~\cite{ref:BeckerDarkDarkBright,ref:AlotaibiDynamicsDarkBright}.\newline
More recently, the addition of quantum fluctuations to the MF approximation was found to explain the stabilization of one-component dipolar Bose gases and two-component Bose-Bose mixtures against collapse~\cite{ref:Kaudau2016,ref:Schmitt2016,ref:PetrovDroplet2015}, sparking new research concerning the liquefaction of Bose gases into self-bound droplets~\cite{ref:PetrovDroplet2015,ref:PetrovLowDim,ref:DropletSemeghini,ref:DropletCabrera} and supersolids~\cite{ref:Chomaz2019,ref:Tanzi2019-1,ref:Tanzi2019-2,ref:MikSupersolid}. In Ref.~\cite{ref:NaidonBubble} it was suggested that for repulsive two-component Bose-Bose mixtures quantum fluctuations likewise may lead to a new phase of matter in between the miscible-immiscible phase transition, a so-called \emph{'mixed bubble'} phase.\newline
As discussed in Ref.~\cite{ref:NaidonBubble} the proposal of this new phase is based on a transformation of the densities $n_{1,2}$ to $n_\pm$ as
\begin{align}
    n_\pm = \frac{\alpha^{-1/2}n_1\pm\alpha^{1/2}n_2}{\sqrt{\alpha+\alpha^{-1}}}\label{eq:nplus},
\end{align}
where $\alpha = \sqrt{g_{22}/g_{11}}$. Along $n_+$ the system's behavior is dominated by the MF contribution, while along $n_-$ the system is sensitive to the usually much weaker quantum fluctuations. This separation of scales allows the grand potential density to predict the onset of new phases along $n_-$, for which it is a convex function if the system is miscible and a concave function if the sytem is immiscible. However, close to the phase transition, where quantum fluctuations become comparable in size to the MF contribution, the grand potential density may be a concave-convex function for $\alpha < 1$ or convex-concave for $\alpha > 1$, allowing for a new phase to emerge. As mentioned above, this new phase has been dubbed a \emph{'mixed bubble'} in a Bose-Bose mixture \cite{ref:NaidonBubble} and can be seen as a pocket of one component trapped within the gaseous medium of the other component. Little is yet known about its properties.\newline
In this Letter, we investigate the formation of these novel \emph{'mixed bubbles'} in a one-dimensional annular confinement close to the miscible-immiscible threshold. We start by exploring the shape of mixed bubbles for a set of population imbalances between the components, probing the predictions made in Ref.~\cite{ref:NaidonBubble} for a uniform, infinite system by varying the criticality parameter $\sqrt{n_+}\delta g/g^{3/2}$. Classifying the miscible-bubble-immiscible phase transition via its collective excitations, a phase (Goldstone) and amplitude (Higgs) mode can be identified. We show that for a binary mixture in a ring, a localized pocket of one component coexists with non-vanishing part of the second component, resulting in a pocket of miscibility within an immiscible mixture. Further, the mixed bubble can support persistent currents in an annular confinement, however exhibiting an avoided level-crossing of rotational states due to a repulsive intercomponent interaction, leading to the analogy of an intrinsic weak link~\cite{ref:RamanathanLink,ref:WrightLink,ref:RyuJosephson}.\newline 
From here on, we follow the notation in Ref.~\cite{ref:NaidonBubble} and signify the miscible phase as $1$$+$$2$, the immsicible phase as $1|2$ and the mixed bubble as $(1$$+$$2)|2$ if $\alpha < 1$ and $(1$$+$$2)|1$ if $\alpha > 1$. Note that according to Ref.~\cite{ref:NaidonBubble} this phase cannot occur for $\alpha = 1$. \newline
\emph{Model} A uniform Bose-Bose mixture in one dimension with equal masses of atoms but potentially unequal short-ranged interactions, including beyond mean-field effects in the Bogoliubov approximation, has energy 
\begin{align}
    E = \sum_{\sigma} E_{\text{kin},\sigma}+\sum_{\sigma,\sigma'}\frac{g_{\sigma\sigma'}}{2}n_{\sigma}n_{\sigma'}+E_{\text{B}}\label{Eq::total_energy},
\end{align}
where $E_\text{B}$ is the Bogoliubov vacuum energy. In one dimension and for homo-nuclear components, $E_\text{B}$ can be written as 
\begin{align}
    -\frac{2}{3\pi}\frac{m^2}{\hbar}\sum_{\pm}c_{\pm}^3\label{Eq::LHYterm}
\end{align}
where the squared Bogoliubov sound velocities are~\cite{ref:PetrovLowDim}
\begin{align}
    c_{\pm}^2=\frac{g_{11}n_{1}+g_{22}n_{2}\pm\sqrt{(g_{11}n_1-g_{22}n_2)^2+4g_{12}^2n_1n_2}}{2m}. \label{eq:BogSoundVel}
\end{align}
The Bogoliubov vacuum energy in Eq.~(\ref{Eq::LHYterm}) is only valid provided the weak-interaction parameter $\eta_{1\text{D}}\approx\sqrt{mg/n}/\hbar\ll 1$. Further, assuming $\delta g \approx \eta_{1\text{D}}$, we set $\delta g = 0$ in the sound velocities $c_{\pm}$ in Eq.~(\ref{eq:BogSoundVel}). The chemical potentials $\mu_{\sigma}$ of the components are then given by $\mu_{\sigma} = \partial E/\partial n_{\sigma}$ and give rise to the respective stationary extended Gross-Pitaevskii equations (eGPe)
\begin{align}
    \begin{split}
        \mu_\sigma\psi_\sigma = \Bigg[-\frac{\hbar^2}{2m}\partial_{\theta\theta}+g_{\sigma\sigma}n_\sigma+g_{\sigma\sigma'}n_{\sigma'}\\
        -\frac{m^{1/2}g_{\sigma\sigma}}{\pi\hbar}\left(\sum_\sigma g_{\sigma\sigma} n_\sigma\right)^{1/2}\Bigg]\psi_\sigma. \label{Eq::GPe}
    \end{split}
\end{align} 
We impose periodic boundary conditions $\psi_{\sigma}(\theta)=\psi_{\sigma}(\theta+2\pi R)$, thus enforcing an annular confinement of circumference $2\pi R$. We set $g = \sqrt{g_{11}g_{22}}$ and use dimensionless units $m=\hbar=R=1$. The resulting stationary eGPe can be found in the supplemental material~\footnote{\label{ref:SuppMat}[url]}.
The order parameter is normalized to the number of particles $N_{\sigma}$ in each component according to $\int_{-\pi}^{\pi}\text{d}\theta|\psi_{\sigma}(\theta)|^2=N_{\sigma}$. The ground state of Eqs.~(\ref{Eq::GPe}) is obtained by imaginary time propagation using a split-step Fourier method. In the later part of this paper we analyze the system under rotation, which requires adding a $-\Omega\hat{L}\psi_{\sigma}$ on the right side of Eqs.~(\ref{Eq::GPe}), with $\hat{L} = -i\partial_\theta$. We also enforce a fixed value of angular momentum $L=L_1+L_2$, where $L_{\sigma}=\int_{-\pi}^{\pi}\text{d}\theta\psi_{\sigma}^*\hat{L}\psi_{\sigma}$ as described in~\cite{ref:Komineas2005,ref:Tengstrand2019}. 
Introducing $N=N_1+N_2$ and $\nu=N_1/N_2$, we illustrate our findings for a two-component system of a mass ratio equal to one and $\alpha = 2.7$, $g=5.0$ and $N=N_1+N_2=10000$. The chosen parameters are proposed to be experimentally reachable configurations to observe the mixed bubble phase in three dimensions for a system of $^{41}$K and $^{39}$K both in hyperfine states $F=1$, $m_{\text{F}}=0$ ~\cite{ref:NaidonBubble,ref:ErricoFeshbach39K,ref:TanziFeshbachKMix}. In order to sample the phase transition we vary $g_{12}$ by changing $\sqrt{n_+}\delta g/g^{3/2}$, using the uniform values for $n_{\sigma}$ in Eq.~(\ref{eq:nplus}).
\newline
\emph{Density distributions.} We first investigate the existence of the mixed bubble phase and the corresponding density distributions for the above choice of parameters. The predicted range in Eqs. (16) and (17) of Ref.~\cite{ref:NaidonBubble} for an infinite, uniform system lies within 
\begin{figure}
    \centering
    \includegraphics[width=0.45\textwidth]{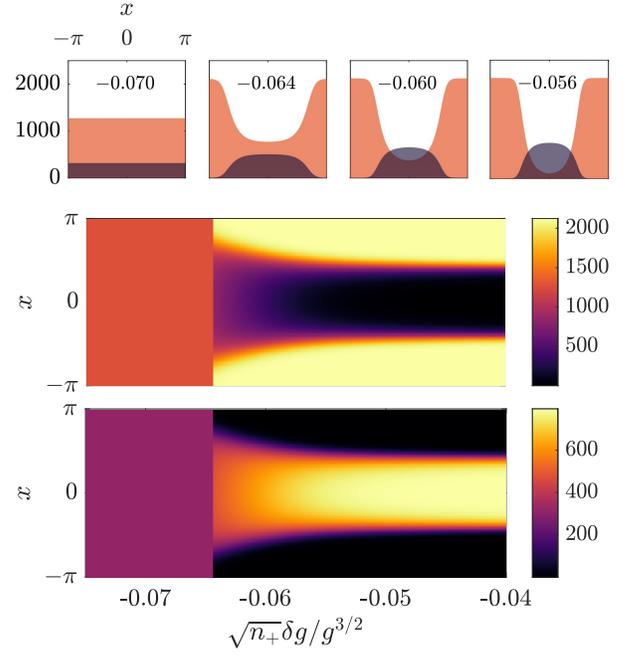}
    \caption{\emph{Top}: Density distributions for four different parameters of $\sqrt{n_+}\delta g/g^{3/2}$ as indicated in the figures, with the orange and black components corresponding to the first and second component respectively. The first distribution shows the system in a miscible state, while the remaining in a mixed bubble state. \emph{Bottom}: Density distributions in the ring for $\nu=4.0$ along the criticality parameter $\sqrt{n_+}\delta g/g^{3/2}$. The top and bottom rows show the first and second component respectively. By increasing $\sqrt{n_+}\delta g/g^{3/2}$ the components start to localize on opposite sides of the ring, resulting in co-existence of a mixed and separated phase until it reaches full phase separation.}
    \label{fig:dens2}
\end{figure}

\begin{align}
\begin{split}
    \frac{\delta g_{\text{min}}}{g^{3/2}}\sqrt{n_+} &= -\frac{1}{4\pi}\frac{(\alpha-1)^2}{\sqrt{\alpha}(\alpha^2+1)^{1/4}},\\
    \delta g_{\text{max}} &= \frac{4(\sqrt{\alpha}+2)}{3(\sqrt{\alpha}+1)^2}\delta g_{\text{min}}\label{Eq::dgminmax}.
\end{split}
\end{align}
Inserting $\alpha=2.7$ then gives the range of $[-0.082,-0.057]$. We continue by choosing $\nu = 4.0$. In Fig.~\ref{fig:dens2} we show examples of density distributions for four values of the criticality parameter, with the first component $|\psi_1|^2$ in orange and the second component $|\psi_2|^2$ in black. As $g_{12}$ increases, the components begin to localize on opposite sides of the ring, shifted by $\pi$. This leads to the predicted co-existence of phase separation and phase mixing induced by quantum fluctuations, eventually resulting in immiscibility by further increasing $g_{12}$. Without the Bogoliubov vacuum energy $E_{\text{B}}$ in Eq.~(\ref{Eq::total_energy}) the system would still be in a fully miscible regime. The full transition is displayed in the bottom panel of Fig.~\ref{fig:dens2}, showing a density plot of the modulation of $\sqrt{n_+}\delta g/g^{3/2}$ along the \emph{x}-axis and the position on the ring on the \emph{y}-axis. Starting from the left side in the miscible phase, the starting point of the bubble phase transition is already significantly larger than the beginning of the interval according to Eqs.~(\ref{Eq::dgminmax}). As one enters the bubble phase, one component localizes on one side, while allowing space for the other component to form the separating part, as depicted in the top panel of Fig.~\ref{fig:dens2}. As one continues to move towards immiscibility, the mixed part of the first component decreases linearly in density with increasing $\sqrt{n_+}\delta g/g^{3/2}$, while the disconnected second component increases in density, reducing its width. Eventually, as the system enters immiscibility, the first component reaches zero density in the overlapping part, eliminating the pocket of miscibility. We find that the predicted range of $(\delta g_{\text{min}},\delta g_{\text{max}})$ in Ref.~\cite{ref:NaidonBubble} is no longer accurate for this trapped Bose-Bose mixture and is highly dependent on $\nu$. This may be due to finite-size effects, compared to the prediction in Ref.~\cite{ref:NaidonBubble} for an infinite system. Indeed, the bubble phase vanishes completely as $\nu\rightarrow1$, however, this process can be reversed for $1/\nu$ if $\alpha\rightarrow1/\alpha$, which results in $(1$+$2)|2\rightarrow(1$+$2)|1$.\newline
The observed bubbles show similarities to dark/bright or magnetic solitons, however, the bubble formation represents a groundstate to the system to bridge the phase transition from phase mixing to separation, while the soliton solutions require $g-g_{12} > 0$ in a description without the Bogoliubov vacuum energy $E_\text{B}$ in Eq.~(\ref{Eq::total_energy})~\cite{ref:BeckerDarkDarkBright}. The dark/bright soliton itself is then induced by imprinting a phase difference onto one component to create a dark soliton~\cite{ref:BurgerDarkSoliton}, filled by the second component, due to the repulsive inter-component interaction $g_{12}$~\cite{ref:BeckerDarkDarkBright,ref:AlotaibiDynamicsDarkBright}. For a magnetic soliton the phase difference is imprinted onto both components leading to a similar density distribution as for the dark/bright soliton~\cite{ref:QuMagneticSoliton,ref:ChaiMagneticSoliton,ref:FarolfiObsMagneticSoliton}.\newline
\begin{figure}
    \centering
    \includegraphics[width=0.45\textwidth]{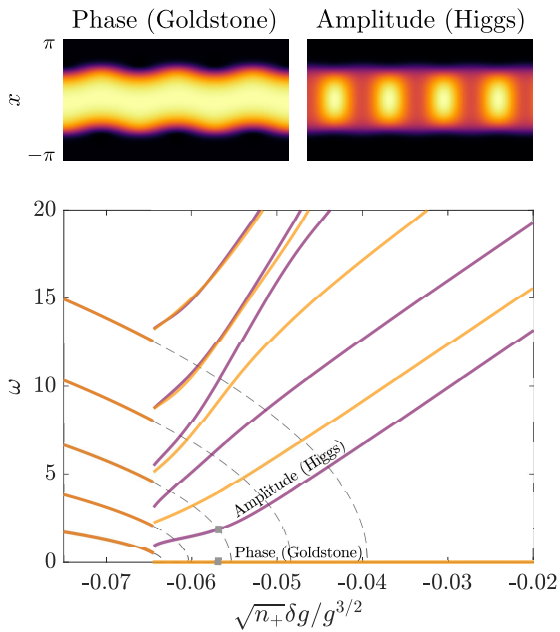}
    \caption{\emph{Top:} Time evolution of the second component of phase (Goldstone) and amptliude (Higgs) mode at $\sqrt{n_+}\delta g/g^{3/2} = -0.0577$ \emph{Bottom:} Collective excitation spectrum of the ten lowest modes for $\nu = 4.0$ across the miscible-bubble-imiscible phase transition from $\sqrt{n_+}\delta g/g^{3/2} = - 0.075$ to $-0.02$. Modes of even parity are displayed in orange and of odd parity in purple. The transition from miscible to bubble occurs at around $-0.0651$. The grey box indicates the position of the Goldstone and amplitude mode shown above. Dashed grey lines indicate the real part of the excitation spectrum of a uniform Bose-Bose mixtur as calculated in the supplemental material}
    \label{fig:spectrum}
\end{figure}
Let us now discuss the collective excitations across the criticality parameter. Applying the standard procedure, we linearize the time-dependent variants of Eqs.~(\ref{Eq::GPe}) to first order around the ground state $\psi_{0,\sigma}$, introducing quasiparticle amplitudes $u_\sigma$ and $v_\sigma$ as 
\begin{align}
    \psi_{\sigma}(\theta,t) = e^{-i\mu_\sigma t}\left[\psi_{0,\sigma}(\theta)+u_\sigma(\theta)e^{-i\omega t}+v^*_\sigma(\theta)e^{i\omega^*t}\right] \label{Eq::coll_linearization},
\end{align}
with $\int_{-\pi}^{\pi}d\theta(|u_{\sigma}(\theta)|^2-|v_{\sigma}(\theta)|^2) = 1$. The resulting equations are given in the supplemental material and the corresponding linear response eigenvalue problem is solved numerically in a real Fourier collocation scheme with a locally optimal block preconditioned four-dimensional conjugate gradient method~\cite{ref:baiLOBP4DCG}. The obtained spectrum is displayed in Fig.~\ref{fig:spectrum}, with even and uneven modes in orange and purple respectively. Starting from the miscible phase, degenerate pairs of uneven and even modes split up upon transitioning from the miscible to the bubble regime. The phase transition is numerically non-continuous in the excitation spectrum due to the external confinement up to a relative stepsize $\approx 10^{-6}$, leading to a jump in excitation frequency. As the bubble localizes, spontaneous breaking of \emph{U}(1) symmetry occurs, leading to a phase (Goldstone) mode in the bubble and immiscible phase. The time evolution of the phase (Goldstone) mode $|\psi_{\sigma}(\theta,t)|^2$, is shown in the top left panel of Fig.~\ref{fig:spectrum}, following 
\begin{align}
    |\psi_{\sigma}(\theta,t)|^2 = |\psi_{\sigma,0}(\theta)|^2 + 2 f_{\sigma}(\theta)\psi_{\sigma,0}(\theta)\cos(\omega t)
\end{align}
with $f_{\sigma}(\theta) = u_{\sigma}(\theta) + v_{\sigma}(\theta)$ ~\cite{ref:HertkornHiggs}. The phase (Goldstone) mode persists through the transition into the immiscible regime. We also identify an amplitude (Higgs) mode in the bubble phase. Note that the densities of both components oscillate in phase at the beginning of the amplitude mode. As $g_{12}$ increases, the amplitude of density oscillation in the first component decreases, until it eventually oscillates out of phase with the second component.\newline

\begin{figure}
    \centering
    \includegraphics[width=0.45\textwidth]{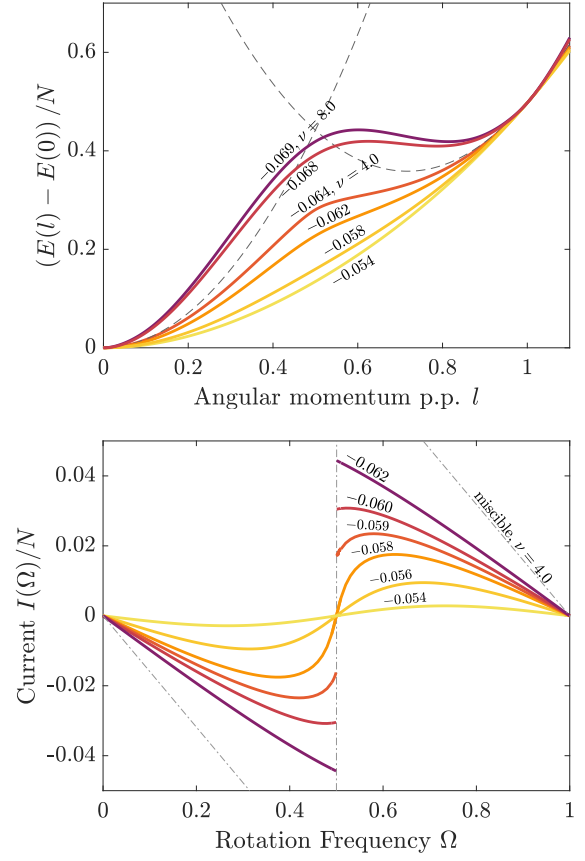}
    \caption{\emph{Top}: Rotational energy per particle in the non-rotating frame at a fixed number of angular momentum per particle $l$. Values of $\sqrt{n_+}\delta g/g^{3/2}$ as given in the plot. Closer to immiscibility, the energy approaches the rigid body limit, while close to miscibility the bubble may support persistent currents for non-zero $l$. \emph{Bottom}: Current $I(\Omega)/N$ through the weak link formed by the bubble. Values of $\sqrt{n_+}\delta g/g^{3/2}$ as given in the plot. Starting with a discontinuity, the current starts bending due to the constriction the bubble provides until the two branches connect at $I(0.5)/N=0$, from which on the current moves from a smeared saw-tooth function to a sinusoidal modulation with increasing intercomponen interaction}
    \label{fig:pers-curr}
\end{figure}

The localization during the phase transition into the mixed bubble suggests to investigate the phase transition by means of the nonrigid rotational-inertia fraction (NRRI)~\cite{ref:LeggetSuperfluidfraction}
$$f_{\sigma} = 1-\lim_{\Omega\rightarrow0}\frac{L_{\sigma}}{N_{\sigma}\Omega}.$$
\begin{figure}
    \centering
    \includegraphics[width=0.45\textwidth]{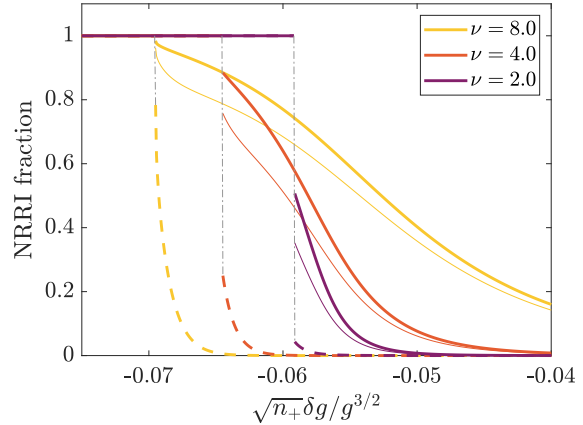}
    \caption{Nonrigid rotational inertia for systems of varying $\nu$ across the miscible-bubble-immiscible phase transition. The second component (dashed line) follows a sharp decline upon localization, turning into a rigid body, while the first component's NRRI (full line) decreases more slowly during the bubble phase. The system's total NRRI as given in the text is shown as a thin full line.}
    \label{fig:NRRI}
\end{figure}
This quantifies the rotational behavior for small rotations by comparing the angular momentum a system picks up under rotation to the angular momentum it would have if it were rigid. As such, a uniform system will have nonrigid rotational inertia equal to unity, and a rigid body equal to zero.\newline 
The NRRI is calculated numerically as a function of $\sqrt{n_+}\delta g/g^{3/2}$ and shown in Fig.\,\ref{fig:NRRI} for three different values of $\nu$. All three systems exhibit a qualitatively similar behaviour, for which the NRRI fraction starts at unity for the mixed phase until criticality. As discussed before, the exact critical value depends on $\nu$, with higher $\nu$ having a lower critical value and coinciding with the onset of localization in the density distributions. For all systems the second component's NRRI (dashed line) approaches zero quickly, signifying the onset of a mixed bubble, only while the first component's NRRI (full line) slowly decreases to zero, signifying the onset of phase separation. The total NRRI of a system is given by the thin lines as $f_\text{s} = (f_{\text{s},1}\nu+f_{\text{s},2})/(1+\nu)$. 
\newline
In the following we will now enforce a certain angular momentum per particle $l$ into systems of variable $\sqrt{n_+}\delta g/g^{3/2}$ and measure its ground state energy per particle in the non-rotating frame as $[E(l)-E(0)]/N$. In the case of a rigid body this leads to a parabola of shape $l^2/2$, while for a uniform system one obtains a concave periodic structure on top of the parabola with minima when a vortex fully nucleated at the center~\cite{ref:BlochSuperfluidity}. We plot this quantity in Fig.~\ref{fig:pers-curr} for the given values of $\sqrt{n_+}\delta g/g^{3/2}$ and find that it follows the trajectory of dampened intersecting parabolas. Evaluating $E(l)$ in dependence of a sytems' NRRI~\cite{ref:MikaelSuperfluid} we show the resulting intersecting parabolas as dashed grey lines in Fig.~\ref{fig:pers-curr} for $\nu=4.0$, $\sqrt{n_+}\delta g/g^{3/2} = -0.064$ with a NRRI fraction of $0.7172$. Here, the dampening effect occurs due to an increased difficulty of the current in component $1$ traversing through the mixed region, until the density in component $1$ in the mixed region goes to zero and a vortex nucleates in the centre of the ring. Despite the dampening effect, the system is still able to support persistent currents for some parameters.
A similar dampening effect is known to occur in atomtronic applications in one-component systems of repulsive Bosons where the rotational symmetry is broken by the addition of a so-called weak link to stir the condensate~\cite{ref:MinguzziWeakLink}. The mixed bubble can here be regarded as an intrinsic weak link with an avoided level-crossing between the rotational states signified by the dampening effect. We calculate the spatially averaged current relative to the rotating ring $I(\Omega) = (2\pi)^{-1}\int_{-\pi}^{\pi}d\theta j(\theta)$ where $j(\theta)=n(\theta)v(\theta)$ is the current density. The velocity $v(\theta)$ in the rotating frame is obtained by inserting $\psi_{\sigma}(\theta) = \phi_{\sigma}(\theta)e^{i\alpha(\theta)}$ into Eqs.~(\ref{Eq::GPe}) and separating the real and imaginary parts. The resulting expression for the gradient of the phase $\alpha'(\theta)$ equals the velocity $v(\theta) = (L-N\Omega)/(2\pi n(\theta))$ ~\cite{ref:MikaelSuperfluid}. The space-averaged current through the weak link in the rotating frame can then be calculated as 
\begin{align}
    I(\Omega) = \frac{1}{2\pi}\int_{-\pi}^{\pi}d\theta j(\theta) = \frac{L-N\Omega}{2\pi}.
\end{align}
The current $I(\Omega)$ is shown in Fig.~\ref{fig:pers-curr} and our description is equivalent to the relation derived in Ref.~\cite{ref:BlochJosephson} and used in Ref.~\cite{ref:MinguzziWeakLink}. Close to the miscible regime, the usual sawtooth behavior of the current obtains a small curvature, but retains its symmetric discontinuity at $\Omega\rightarrow0.5$. The mixed bubble grows in size with increasing intercomponent interaction strength and provides a stronger obstacle to overcome for the flow in component one. Thus, the currents' branches start bending towards zero as $\Omega\rightarrow0.5$ until they eventually connect and become continuous with $I(0.5)=0$. The current then resembles the behavior of a one-component system with a strong weak link~\cite{ref:MinguzziWeakLink}. Initially this is displayed by a dampened sawtooth function, which moves towards a sinuisoidal oscillation as the system approaches immiscibility.\newline
Similarly to self-bound droplets, quantum fluctuations originating from the Bogoliubov vacuum energy in Eq.~(\ref{Eq::LHYterm}) introduce a new phase of Bose-Bose mixtures inbetween the miscible-immiscible phase transition confirming the recent prediction made in Ref.~\cite{ref:NaidonBubble}. We found that intriguingly the \emph{'mixed bubbles'} act like a single component trapped in a gaseous medium as a pocket of miscibility exhibiting non-trivial non-rigid rotational inertia. When enforcing a certain amount of angular momentum on the system, the mixed bubble mimics a weak link, with an avoided level crossing between the consecutive rotational states due to the repulsive interspecies interactions. However, the system is still able to support persistent currents around the ring for some parameters, as the reduction of the cusp leads to a plateau. Our work, here restricted to an investigation of the newly discovered mixed bubble phase, opens many new questions. It would be interesting to explore bubble dynamics and to extend the analysis also to the two- and three-dimensional case, in a non-annular trap and for a heavily mass-imbalanced system such as a $^{174}\text{Yb-}^{7}\text{Li}$ mixture. Particularly interesting perspectives emerge for atomtronics applications (see the recent review~\cite{ref:revAmico}), where the weak link dynamics play a crucial role.
\bigskip
\begin{acknowledgments}
{\it Acknowledgements.} This work was financially supported by the Knut and Alice Wallenberg Foundation, the Swedish Research Council and NanoLund. We thank J. Bengtsson and J. Bjerlin for fruitful discussions about the excitation spectrum and A. Recati for his suggestions on the solitonic structure.
\end{acknowledgments}

\bibliography{lib.bib}

\pagebreak
\begin{center}
\widetext
\textbf{\large Supplemental Material: Mixed Bubbles in a one-dimensional Bose-Bose mixture}

\end{center}

\begin{center}
    P. Stürmer, M. Nilsson Tengstrand, and S.M.Reimann \\
    \emph{Mathematical Physics and NanoLund, Lund University, Box 118, 22100 Lund, Sweden}\\
    \date{\today}
\end{center}
\setcounter{equation}{0}
\setcounter{figure}{0}
\setcounter{page}{1}
\makeatletter

\section{Unitless Gross-Pitaevskii equations}
A one-dimensional two-component Bose-Bose mixture is described by the GPe in Eq.~\ref{Eq::GPe}. In the maintext we mentioned the scaling variances $\hbar = m = R = 1$, as well as $\alpha = \sqrt{g_{22}/g_{11}}$ and $g = \sqrt{g_{11}g_{22}}$. This results in the following coupled system of time-independent GPe:
\begin{align}
\begin{split}
        \mu_1\psi_1 &= \left[-\frac{1}{2}\partial_{xx} +\frac{g}{\alpha}n_1+g_{12}n_2-\frac{g^{3/2}}{\pi\alpha}\left(\frac{n_1}{\alpha}+n_2\alpha\right)^{1/2}\right]\psi_1, \\
        \mu_2\psi_2 &= \left[-\frac{1}{2}\partial_{xx}+g\alpha n_2+g_{12}n_1-\frac{g^{3/2}\alpha}{\pi}\left(\frac{n_1}{\alpha}+n_2\alpha\right)^{1/2}\right]\psi_2.
\end{split}
\end{align}

\section{Bogoliubov-de Gennes equations}
The study of collective excitations results from the linearsiation process described in the main text. For two components this results in a $4N\times4N$ linear response eigenvalue problem $\textbf{Mv} = \omega \textbf{v}$, where $N$ is the dimension of each block matrix and $\mathbf{v} = (u_1(x),v_1(x),u_2(x),v_2(x))^\top$:
\begin{align}\mathbf{M} = 
\begin{bmatrix}
X_{12} & Y_1 & Z & Z\\
-Y_1 & -X_{12} & -Z & -Z\\
Z & Z & X_{21} & Y_2\\
-Z & -Z & -Y_2 & -X_{12}
\end{bmatrix}
\end{align}
with 
\begin{align}
    \begin{split}
        X_{\sigma\sigma'} &= -\frac{1}{2}\partial_{xx}+2g_{\sigma\sigma}n_{\sigma} +g_{\sigma\sigma'}n_{\sigma'}-\frac{g_{\sigma\sigma}}{2\pi}\frac{3g_{\sigma\sigma}n_{\sigma}+2g_{\sigma'\sigma'}n_{\sigma'}}{\left(g_{11}n_1+g_{22}n_2\right)^{1/2}}-\mu_{\sigma}\\
        Y_{\sigma} &= g_{\sigma\sigma}n_{\sigma}-\frac{g_{\sigma\sigma}}{2\pi}\frac{g_{\sigma\sigma}n_{\sigma}}{\left(g_{11}n_1+g_{22}n_2\right)^{1/2}} \\
        Z &= g_{12}\sqrt{n_1n_2}-\frac{g}{2\pi}\frac{g\sqrt{n_1n_2}}{\left(g_{11}n_1+g_{22}n_2\right)^{1/2}}.
    \end{split}
\end{align}
Let us now switch the second and the third column of $\textbf{M}$ and perform a orthogonal similarity transformation $\mathbf{J}^\top\mathbf{M}\mathbf{J}$, where
\begin{align}
    \mathbf{J} = \frac{1}{\sqrt{2}}
    \begin{bmatrix}
    I& I \\
    I & -I
    \end{bmatrix},
\end{align}
with $I$ the $2N\times2N$ identity matrix. Further we introduce $\mathbf{K},\mathbf{M}$ as
\begin{align}
    \mathbf{K} = 
    \begin{bmatrix}
    Y_1+X_{12} & 0 \\
    0 & Y_2+X_{21}
    \end{bmatrix},
    \mathbf{M} = 
    \begin{bmatrix}
    X_{12}+Y_1 & 2Z \\
    2Z & X_{21}+Y_2
    \end{bmatrix},
\end{align}
such that 
\begin{align}
    \mathbf{J^\top MJ} =\tilde{\mathbf{M}}=
    \begin{bmatrix}
    0 & K \\
    M & 0
    \end{bmatrix},
\end{align}
which is the common shape for a linear response eigenvalue problem~\cite{ref:bai2014minimization}. The new eigenvectors $\mathbf{a}=(a_1(x),a_2(x),b_1(x),b_2(x))^\top$ then come out as $\mathbf{a}=\mathbf{J}^\top\mathbf{v}$, such that the full eigenvalue problem takes the form of
\begin{align}
    \tilde{\mathbf{M}}\mathbf{a} = \omega\mathbf{a}
\end{align}
\section{Excitation spectrum of uniform two-component system}
The excitation spectrum for uniform densities of $n_1=N_1/2\pi$ and $n_2=N_2/2\pi$ can be readily calculated by expanding $u(x)$ and $v(x)$ in plane waves as~\cite{ref:bookPethickSmith} $$\mathbf{v}(x) = \sum_{p=-\infty}^{\infty}\frac{\text{e}^{ipx}}{\sqrt{2\pi}}\tilde{\mathbf{v}}_p.$$ Substituting this into $\mathbf{Mv}=\omega\mathbf{v}$ gives the frequencies as:
\begin{align}
\begin{split}
    \omega_{p,\pm}^2 &= \frac{p^2}{4\pi}\Bigg[p^2\pi+\sum_{\sigma,\sigma'}g_{\sigma\sigma}N_{\sigma}(1-g_{\sigma\sigma}\eta)\pm\sqrt{\left(g_{11}N_1(1-g_{11}\eta)-g_{22}N_2(1-g_{22}\eta)\right)^{2}+4N_1N_2(g_{12}-g_{11}g_{22}\eta)^2}\Bigg]
    \end{split}
\end{align}
with $\eta = \left[2\pi(g_{11}N_1+g_{22}N_2)^{1/2}\right]^{-1}$, which is shown as light-grey dashed lines in Fig.~\ref{fig:spectrum}.
\end{document}